\documentclass[prl,aps,amsmath,amssymb,twocolumn,showpacs]{revtex4}
\usepackage{graphicx}
\usepackage{color}
\usepackage{amsmath}
\usepackage{amssymb}
\newcommand{\be}{\begin{equation}}
\newcommand{\ee}{\end{equation}}
\newcommand{\bd}{\begin{displaystyle}}
\newcommand{\ed}{\end{displaystyle}}
\newcommand{\ba}{\begin{eqnarray}}
\newcommand{\ea}{\end{eqnarray}}
\DeclareMathSymbol{\lang}{\mathord}{symbols}{"68}
\DeclareMathSymbol{\rang}{\mathord}{symbols}{"69}
\DeclareMathSymbol{\openbra}{\mathord}{symbols}{"68}
\DeclareMathSymbol{\closeket}{\mathord}{symbols}{"69}

\newcommand{\aver}[1]{{\lang #1 \rang}}
\begin{document}
\title{Manifestation of Hamiltonian chaos in an open quantum system with ballistic atoms in an
optical lattice}
\author{V.Yu. Argonov and S.V. Prants}
\affiliation{Laboratory of Nonlinear Dynamical Systems,\\
V.I.Il'ichev Pacific Oceanological Institute
of the Russian Academy of Sciences,\\ 43 Baltiiskaya st.,
690041 Vladivostok, Russia}
\begin{abstract}
 Manifestation of dynamical 
instability and Hamiltonian chaos in the fundamental near-resonant 
matter-radiation interaction has been found analitically and in a Monte Carlo simulation in the behavior of atoms moving 
in a rigid optical lattice. Character of diffusion of spontaneously emitting atoms 
changes abruptly in the range of the values of parameters and initial conditions 
where their Hamiltonian dynamics is shown to be chaotic.    
\end{abstract}
\maketitle

 Atoms in an optical lattice, formed by a laser standing
wave, is an ideal system for studying
quantum nonlinear dynamics. Operating at low temperatures and 
 controlling the lattice parameters, experimentalists
 now are able to tailor practically one-dimensional potentials
 and manipulate with internal and external degrees of freedom of atoms. 
Experimental study of quantum chaos
has been carried out with ultracold atoms interacting with a periodically modulated
optical lattice \cite{MR94}.
To suppress spontaneous emission and provide Hamiltonian quantum dynamics atoms are detuned far
from the optical resonance. Adiabatic elimination of
the excited state amplitude leads to an effective
Hamiltonian for the external motion \cite{GSZ92} 
corresponding to a 3/2 degree-of-freedom classical
system which has a mixed phase space with regular islands embedded in 
a chaotic sea.
De Brogile waves of ultracold atoms have been shown to
demonstrate under appropriate conditions the effect
of dynamical localization which means quantum suppression of chaotic diffusion
\cite{GSZ92,MR94}. Decoherence due to spontaneous emission tends to suppress 
this  quantum effect and restore classical-like dynamics  \cite{KO98}.

A new arena of quantum nonlinear dynamics with atoms in optical
lattices is opened if we work near the optical resonance and take the 
internal dynamics into account. In the Hamiltonian approximation, when 
one neglects spontaneous emission (SE), the coupling
of internal and external atomic degrees of freedom
has been shown to produce a number of nonlinear effects in rigid 
(i.e. without any modulation) optical lattices: chaotic Rabi
oscillations, chaotic atomic transport, dynamical fractals, and 
L\'evy flights \cite{PRA01,PRA07}.
In real life the dynamics of atoms in near-resonant laser fields is not
deterministic because of SE.
 The problem of interrelation between deterministic chaos and noise
 is rather  general. Natural systems are
 subject to noise which, usually, acts continuously. If
noise is practically continuous and comparatively weak we can study
in which way it affects  chaotic deterministic evolution of the
system under consideration.
Spontaneous emission is a kind of a shot noise which is not small because SE 
recoils may change the internal state significantly.
In this Letter we demonstrate analitically and in a  Monte Carlo 
simulation that manifestation of dynamical 
instability and Hamiltonian chaos in the fundamental near-resonant 
matter-radiation interaction can be found in the behavior of atoms moving 
in a rigid optical lattice.

Since we study manifestation of quantum nonlinear effects in ballistic transport
of atoms, when the average atomic momentum is very large as compared
with the photon momentum $\hbar k_f$, the translational motion is described classically by 
Hamilton equations. We start with the Hamilton-Schr\"odinger
equations of motion for a two-level atom in a standing light wave which
have been derived in Refs. \cite{Acta06,epl}: 
\begin{equation}
\begin{aligned}
\dot x&=\omega_r p,\quad
\dot p=-u\sin x+\sum\limits_{j=1}^{\infty}p_j\delta(\tau-\tau_j),
\\
\dot u&=\Delta v+\frac{\gamma}{2}uz-u\sum\limits_{j=1}^{\infty}\delta(\tau-\tau_j),
\\
\dot v&=-\Delta u+2 z\cos x+\frac{\gamma}{2}vz-v\sum\limits_{j=1}^{\infty}\delta(\tau-\tau_j),
\\                                         
\dot z&=-2 v\cos x-\frac{\gamma}{2}(u^2+v^2)-(z+1)\sum\limits_{j=1}^{\infty}\delta(\tau-\tau_j),
\end{aligned}
\label{mainsys}
\end{equation}
where $x\equiv k_f X$ and $p\equiv P/\hbar k_f$ are 
normalized atomic center-of-mass position and momentum, 
$u$, $v$, and $z$ are synphase and quadrature
components of the atomic electric dipole moment
and the population inversion, respectively. The length of the Bloch vector, $u^2+v^2+z^2=1$, is conserved.
The dot denotes differentiation with respect to
the normalized time $\tau\equiv \Omega t$. The {values of the} 
normalized
decay rate $\gamma\equiv \Gamma/\Omega$ and the recoil frequency
$\omega_r\equiv\hbar k_f^2/m_a\Omega$ are chosen to be
$\gamma=3.3\cdot 10^{-3}$ and $\omega_r=10^{-5}$ and correspond to
a cesium atom ($\lambda_a=852.1$ nm
and $\Gamma=3.2\cdot10^7$ s$^{-1}$) in a strong
field with the Rabi frequency $\Omega=10^{10}$~s$^{-1}$. 
So, the normalized detuning between the field and atomic frequencies, $\Delta\equiv(\omega_f-\omega_a)/\Omega$
, is a single variable parameter. In Eqs. (\ref{mainsys}) $\tau_j$ are random
time moments of SE events
and $p_j$ are random recoil momenta with the
values between $\pm 1$ (1D case).
In terms of the normalized time $\tau$ the mean
frequency of SE events is 
$\gamma (z+1)/2$. At  $\tau=\tau_j$, the
atomic variables change as follows: $p\to p+ p_j$,
$u\to 0,\ v\to 0,\ z\to -1$.

Equations (\ref{mainsys}) with $\gamma=0$ and without the terms containing 
delta-functions describe Hamiltonian coherent evolution
of the internal and external degrees of freedom of an atom
that has been shown \cite{PRA01} to be chaotic (in the sense of exponential
sensitivity to small changes in initial conditions) in  certain ranges
of values of the parameters $\omega_r$ and $\Delta$ and initial momenta. 
With comparatively small values
of the initial atomic momentum $p_0$, atoms may wander in
an optical lattice with alternating trapping in the wells of
the optical potential and flights over its hills. It is a
kind of a random walking that may occur without any modulation
of the lattice parameters and/or any noise like SE \cite{PRA01,PRA07}.

In this work we consider only fast ballistic atoms
which never change the direction of motion. There is a range of large values of initial momentum $p_0$ where the
maximal Lyapunov exponent $\lambda$ of the Hamiltonian
equations of motion has been computed to be
positive \cite{PRA07}.
It means that the momentum of a ballistic atom without SE may oscillate
in a deterministic but chaotic way around a mean value $\aver{p}$.
The central question of the present study is the following. 
In which way the Lyapunov instability 
and Hamiltonian chaos, that may occur between SE events,
manifest itself in ballistic atomic transport which is a stochastic process 
due to SE?

\begin{figure*}[htb]
\includegraphics[width=0.96\textwidth,clip]{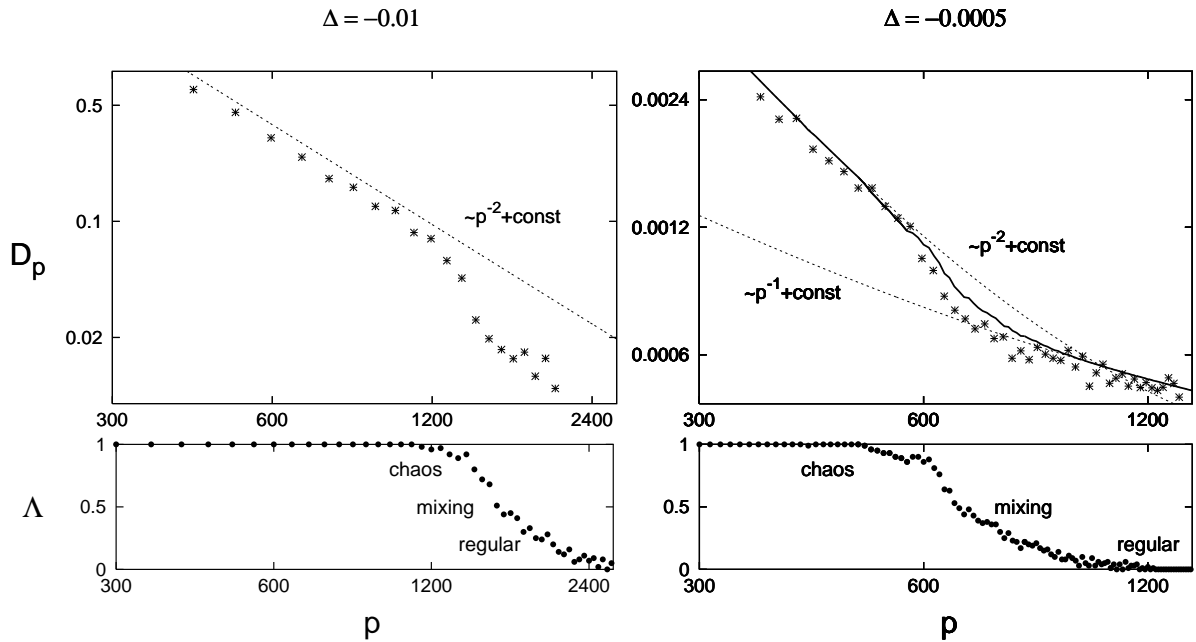}
\caption{Correlation between the average momentum diffusion coefficient 
in a log-log scale (in units of $\hbar^2k^2_f\Omega$) and probability of 
Hamiltonian chaos $\Lambda$ in
their dependencies on atomic momentum $p$ (in units of $\hbar k_f$) 
at $\Delta=-0.01$ and $\Delta=-0.0005$.
Dashed lines with the slope $~p^{-2} + {\rm const}$ and $~p^{-1} + {\rm const}$ 
are theoretical curves (\ref{Dpc}), (\ref{D_reg}), valid in the regimes of 
Hamiltonian chaos and order, respectively. Solid line is a theoretical curve 
(\ref{DprDpc}) derived to fit the numerical dependence. An abrupt change in the decay laws for 
$D_p$  occurs for those values of $p$ where the transition from order to 
chaos takes place in the Hamiltonian dynamics.}
\label{fig1}
\end{figure*}

To answer the question we simulate Eqs. (\ref{mainsys}) by a Monte Carlo method (for details see 
\cite{Acta06}) and compute atomic trajectories in the momentum space
 to find the momentum diffusion
coefficient $D_p$ as a function  of the momentum $p$.     
The results are compared with the maximal Lyapunov exponent $\lambda$
computed with the Hamiltonian anologue of the set (\ref{mainsys} (without SE). More
exactly, as a measure of Hamiltonian chaos, we compute {\it chaos probability} $\Lambda\equiv\aver{2\theta(\lambda)-1}$, where $\theta(\lambda)$ is
a Heaviside function, which is equal to $0$ for $\lambda<0$, $1/2$ for $\lambda=0$, and $1$ for $\lambda>0$.
 The values of $\Lambda$ in Fig.~\ref{fig1} have been computed by averaging
over many atomic trajectories with close values of $p$. If $\lambda >0$ with 
all those atoms, then $\Lambda=1$, and we have
Hamiltonian chaos with probability $1$. If $\lambda=0$
then $\Lambda=0$, and
 the motion is regular with the probability  $1$. The values in the range
 $0<\Lambda<1$ mean that chaotic and regular trajectories
are mixed in a small range of values of initial
momenta, and chaos probability is proportional to the fraction of atoms with positive $\lambda$s.
Fig.~\ref{fig1} demonstrates a correlation between the regimes of chaotic 
(regular) Hamiltonian transport and
the behavior of the momentum diffusion coefficient 
$D_p$ for spontaneously emitting atoms. Beginning with those values of the momentum
$p$, for which the probability of
Hamiltonian chaos becomes smaller than $1$ (see the  lower panels in
Fig.~\ref{fig1}), one observes an abrupt transition
to a more regular regime of motion with another law of decay of $D_p$ 
(see the upper panels in Fig.~\ref{fig1}).

 We stress that the
atomic transport in reality is stochastic with all
the values of $p$ due to SE, but the measure of its stochasticity, $D_p$,
 decays rapidly in the same range of the momenta where the
 Hamiltonian analogue of the system demonstrates a transition from
 chaos to order. It is more
important that this difference could be measured in real experiments and
would provide us with direct signatures of
atomic Hamiltonian chaos in terms of transport characteristics
which are more easy to measure than the Rabi oscillations.

In what follows we estimate the diffusion coefficient $D_p$ 
when the corresponding Hamiltonian ballistic transport is chaotic and regular. 
In the weak Raman-Nath approximation,
$\omega_rp^2/2\gg|u\cos x+\Delta z/2|$, when the atomic
kinetic energy is not strictly a constant, but  
much larger than the potential one, the momentum fluctuations between SE are small. 
In Ref. \cite{epl} we have shown that at small detunings, $|\Delta|\ll 1$,
the evolution of the total energy $H\equiv\omega_rp^2/2-u\cos x-\Delta z/2
$ (which is a constant in the absence of dissipation) is a quasi-random
 process with sudden changes in $H$, when SE occurs, and a slight linear drift in between. 
 We can treat the evolution of $H$ as the following mapping: 
\be
\begin{aligned}
&H_j=H_{j-1}+\omega_r p(\tau_j) p_j+\frac{\omega_r}{2} p_j^2+\frac{\Delta}{2}+\\
+&u(\tau_j)\cos x(\tau_j)+\frac{\Delta}{2}z(\tau_j)+\frac{\Delta\gamma}{4}\aver{1-z^2}(\tau_j-\tau_{j-1}),
\label{DH}
\end{aligned}
\end{equation}
where $H_j$ is a value of the energy just after $j$-th SE,  
$x(\tau_j)$, $u(\tau_j)$, $z(\tau_j)$, and $p(\tau_j)$ are values  
of the corresponding variables just before $j$-th SE 
 which are determined by the evolution between SE events. 
 The last term with the averaging over a time exceeding the period
of the Rabi oscillations is a result of an energy drift between SE events.
In general, this random walk is asymmetric. There is  a friction force $F\equiv\aver{\dot p}$ which
can accelerate or decelerate  atoms in average. 
The measure of momentum fluctuations
in an atomic flight (duration of which exceeds largely 
$\aver{\tau_j-\tau_{j-1}}$) is a momentum diffusion coefficient $D_p$ 
which can be written as 
\be
\begin{aligned}
\ D_p\simeq\frac{\aver{(H_j-H_{j-1})^2}-\aver{H_j-H_{j-1}}^2}{2\omega_r^2 p^2\aver{\tau_j
-\tau_{j-1}}},
\end{aligned}
\ee
 where the average value of the momentum in the weak Raman-Nath approximation 
is $p\simeq\sqrt{2H/\omega_r}$. 
Using  the largest second and fifth terms in Eq. (\ref{DH}), 
we can estimate $D_p$ as follows:
\be
\begin{aligned}
D_p\simeq\frac{\gamma}{{12}}+\frac{\aver{u^2(\tau_j)}\gamma}{8\omega_r^2 p^2}.
\label{D2}
 \end{aligned}
\ee
 All the other terms in Eq.(\ref{DH}) are small since $|\Delta|\ll 1$, $|z| 
\sim 1$, and
$|u|\gg|\Delta|$. In deriving Eq. (\ref{D2}), we put 
$\aver{u\cos x}\simeq0$, $\aver{\tau_j-\tau_{j-1}}\simeq2/\gamma$,
and $\aver{p_j^2}=1/3$. 

To estimate the value of $\aver{u^2(\tau_j)}$ in Eq. (\ref{D2}) we use
 the results of our theory \cite{PRA07} where we have shown, 
that in the absence of SE the variable $u$ can be approximated by a constant 
when atoms move between nodes of a standing wave (it fact, it performs 
shallow oscillations) which changes suddenly its value when they cross any node at $\cos x=0$.  
Spontaneous emission results 
in additional  jumps, $u\to 0$, but between SE events one can
 approximately describe the dynamics using the Hamiltonian theory.
In the chaotic case, the evolution of $u$ can be approximated as
a stochastic mapping  
\be
\begin{aligned}
u_m\simeq|\Delta|\sqrt\frac{\pi}{\omega_r p}\sin\phi_m+u_{m-1}, 
\end{aligned}
\ee
where $u_m$ are  values of $u$ after $m$-th node crossing (starting with 
 the latest SE event), 
$\phi_m$ are random phases in the range $[0, \pi]$. The index $m$ 
 increases by $1$ just after each node crossing and jumps to zero just after  
SE event. This map is obtained from the expression (11) in Ref. \cite{PRA07} 
in the limit $|u|\ll 1$ which
is valid because we have $u=0$ after any act of SE , and the values of $u$ 
never go far away from zero due
to small magnitudes of jumps. Between the acts of SE, a sequence of values of $u$
 looks like a Markov chain of random jumps where the next state depends only 
on the previous one.
In the weak Raman-Nath approximation, the number of node crossings between SE   
can be estimated in the average to be $\aver{M}=2\omega_r p/(\gamma\pi)$. 
Now we can estimate the value of $u(\tau_j) \simeq u_M$ and, using Eq.(\ref{D2}), 
get the following 
formula for the momentum diffusion  coefficient in the regime of Hamiltonian chaos
\be
\begin{aligned}
D_{ch}\simeq&\frac{\gamma}{{12}}+\frac{\Delta^2}{8\omega_r^2 p^2}.
\label{Dpc}
\end{aligned}
\ee
In Fig.~\ref{fig1} this function (\ref{Dpc}) is shown by the dashed lines in 
a log-log scale. 
It fits well numerical data in the range of atomic momenta where Hamiltonian
dynamics is fully chaotic, i.e. at $\Lambda=1$. The formula (\ref{Dpc}) is 
valid in a wide range of moderately small detunings, but it does not work 
in the Hamiltonian mixing and regular regimes.

At very small detunings, we may estimate $D_p$
both in the fully chaotic ($\Lambda=1$) and regular ($\Lambda=0$) regimes. 
With fast atoms at $|\Delta|\lll 1$,
we can use the strong Raman-Nath approximation (neglecting the
momentum fluctuations between SE at all) and adopt the simple linear 
law of motion $x=\omega_r p\tau$. In Ref. \cite{PRA07} we have derived 
a formula (see Eq. (A3) therein) for the value of $u$ after crossing the first 
node. So, we get for $u$ after crossing the m-th node
\be
\begin{aligned}
u_m\simeq\Delta 
 \left[\sqrt\frac{\pi}{\omega_r p}\left(v_0\cos\left(\frac{2}
{\omega_r p}-\frac{\pi}{4}\right)
+(-1)^m\right.\right.\\\times z_0\left.\left.\sin\left(\frac{2}
{\omega_r p}-\frac{\pi}{4}\right)\right)+(-1)^m z_0\right]+u_{m-1} ,
\end{aligned}
\ee
 where $v_0$ and $z_0$ are constant values of $v$ and $z$ at $x=\pi k$,
$k=1,2, \dots$. Now jumps of $u$ are not random, and in the range between 
two SE events a trajectory
looks like a ladder with odd and even jumps of different size
with the total number of steps $M$. It can be shown that 
$u(\tau_j)\simeq  u_{M}\simeq M\Delta   v_0  
\sqrt{\pi/\omega_r p}\cos({2/\omega_r p-\pi/4}).$
Since $v_0$ differs after different SE events, we put $\aver{v_0^2}\simeq 1/2$.
At very small detunings,
the  diffusion coefficient, corresponding to Hamiltonian regular motion 
($\Lambda=0$), is
\be
D_{reg}\simeq\frac{\gamma}
{{12}}+\frac{  \Delta^2}{ 4\gamma\omega_rp \pi  }\cos^2\left(\frac{2}{\omega_r p}-\frac{\pi}{4}\right)\simeq\frac{\gamma}{{12}}+\frac{  \Delta^2  }{ 8\gamma\omega_rp \pi  }.
\label{D_reg}
\ee
Thus, we have the analytic expressions for $D_p$
 in the regimes of Hamiltonian  chaos ($\Lambda=1$) and Hamiltonian order 
($\Lambda=0$).
In general case, $0\leq\Lambda\leq1$, we 
suppose a linear law for the momentum diffusion:
\be
\begin{aligned}
D_p\simeq(1-\Lambda)D_{reg}+\Lambda D_{ch}\simeq\\\simeq\frac{\gamma}
{{12}}+\frac{\Delta^2}{8\omega_rp}\left(  \frac{  1-\Lambda }{ \gamma \pi  }+\frac{\Lambda}{\omega_rp}\right)\label{DprDpc}.
\end{aligned}
\ee
This function is shown by the 
solid line in the upper right panel in Fig.~\ref{fig1}.

Let us consider a small cloud of atomic gas moving in one direction 
with the mean momentum $\aver{p}$. Initial position and momentum distributions are supposed to be 
Gaussian with the standard deviations $\sigma^2_x\equiv\aver{(x-\aver{x})^2}$ and  
 $\sigma^2_p\equiv\aver{(p-\aver{p})^2}$. The momentum diffusion coefficient is 
  $D_p=d(\sigma^2_p)/(2d\tau)$.
 The temperature of atomic gas and its heating speed
(in units K/s) are
\begin{equation}
T\equiv\frac{2\aver{E_k}}{k_B}=\frac{\hbar^2k^2_f\sigma^2_p}{m_ak_B},
\quad\frac{dT}{dt}=\frac{2\hbar^2k^2_f\Omega D_p}{m_ak_B},
\end{equation}
 where $E_k$ is the atomic kinetic energy (in J)
in the center-of-mass moving frame.  The heating speed is proportional to
$D_p$ which has been shown in this Letter to demonstrate different 
behavior in the regimes of regular and chaotic Hamiltonian dynamics. 

In real experiments the measurable quantity is a linear cloud size 
$L\equiv2\sigma_x/k_f$ (in meters).
At small observation times $\tau\ll p/F$ and small temperatures
$\sigma_p\ll\aver{p}$,
we can approximate $D_p$ by a constant for all the atoms in a cloud  
which does not change significantly under the action of 
the force $F$. In this approximation we find: $\sigma^2_x=\sigma^2_x(0) + 
(1/2)\omega^2_r\sigma^2_p(0)\tau^2+(2/3)D_p\omega^2_r\tau^3$.  
We have computed $L$ with Eqs. (\ref{mainsys}) and with that formula 
and found a good correlation between the results and a strict difference 
in extension of atomic clouds in the presence 
and absence of Hamiltonian chaos.

In conclusion, we have found in numerical experiments the 
manifestation of dynamical instability and Hamiltonian chaos
in ballistic motion of two-level atoms in a near-resonance
 standing-wave field. The effect
of dynamical chaos in the fundamental atom-light
interaction is masked by random events of SE. Nevertheless,
we proved analytically and numerically that, under certain conditions, 
there exists 
 a clear correlation between the behavior of the momentum
 diffusion coefficient $D_p$ and chaos probability $\Lambda$.
To detect and quantify this effect in a real experiment, we propose
to measure linear extensions $L$ of atomic clouds with different values
of the mean atomic momentum $\aver{p}$. We predict that beginning with those
values of $\aver{p}$, for which Hamiltonian chaos probability becomes to 
be $1$, the value of $L$ for the corresponding atomic clouds should increase sharply.

This work was supported  by the Russian Foundation for Basic Research
(project no. 06-02-16421) and by the Presidential grant no. MK -- 1680.2007.2.

\end{document}